\documentclass[twocolumn,prl,reprint, amsmath,amssymb,showpacs,superscriptaddress]{revtex4-1}

\usepackage{graphicx}
\usepackage{dcolumn}
\usepackage{bm}
\usepackage{graphicx}
\usepackage{epsfig}
\usepackage{epsf}
\usepackage{amssymb}
\usepackage{amsmath}
\usepackage{amsthm}
\usepackage{multirow}
\usepackage{cases}
\usepackage[colorlinks=true,linkcolor=blue,citecolor=blue,pdfauthor={ },pdftitle={ },pdfsubject={ },pdfkeywords={ }]{hyperref}
\usepackage{pgfplots}
\usepackage{lipsum}

\begin{document}

\title{Hybrid Quantum-Classical Approach to  Quantum Optimal Control}

\author{Jun Li}
\email{lijunwu@mail.ustc.edu.cn}
\affiliation{Beijing Computational Science Research Center, Beijing 100193, China}

\author{Xiaodong Yang}
\affiliation{Hefei National Laboratory for Physical Sciences at Microscale and Department of Modern Physics, University of Science and Technology of China, Hefei, Anhui 230026, China}

\author{Xinhua Peng}
\email{xhpeng@ustc.edu.cn}
\affiliation{Hefei National Laboratory for Physical Sciences at Microscale and Department of Modern Physics, University of Science and Technology of China, Hefei, Anhui 230026, China}
\affiliation{Synergetic Innovation Centre of Quantum Information $\&$ Quantum Physics,
University of Science and Technology of China, Hefei, Anhui 230026, China}

\author{Chang-Pu Sun}
\affiliation{Beijing Computational Science Research Center, Beijing 100193, China}

\begin{abstract} 
A central challenge in quantum computing is to identify more computational problems for which utilization of  quantum resources can offer significant speedup. Here, we propose  a  hybrid quantum-classical   scheme to tackle the quantum optimal control problem.  We show that the most computationally demanding part of   gradient-based   algorithms, namely computing the fitness function and its gradient for  a control input, can be accomplished by the process of  evolution and measurement on a quantum simulator.    By posing queries to and receiving messages from the quantum simulator, classical computing devices update the control parameters  until an optimal control solution is found. To  demonstrate the quantum-classical scheme in experiment, we use a nine-spin nuclear magnetic resonance  system, on which we have  succeeded in preparing a seven-correlated quantum state without involving  classical computation of the large Hilbert space evolution. 
\end{abstract}


\maketitle

Quantum computing promises to deliver a new level of computation power \cite{NC00}. Enormous efforts have been made in exploring the possible ways of using quantum resources to speed up computation. While the fabrication of  a full-scale universal  quantum computer   remains a huge technical challenge \cite{LJLNMO10},  special-purpose quantum simulation can be an alternative  \cite{CD10,CZ12,HCTDL12}. 
Quantum simulators are designed to imitate specific quantum systems of interest, and are expected to provide  significant speed-up over their classical counterparts \cite{Feynman82}. Quantum simulation has found important applications for a great variety of computational tasks, such as solving linear equations \cite{HHL09,C09}, simulating condensed-matter systems \cite{ASK15}, calculating  molecular properties \cite{Lanyon10,Malley16} and certificating     untrusted quantum devices  \cite{WGFC14}.  
However, in view of experimental implementation, most of the proposed algorithms have hardware requirements still far beyond the  capability of near-term quantum devices.

Recent advances towards building   a modest-sized quantum computer have led to   emerging interest in a  quantum-classical hybrid approach \cite{BWMHT16,BSS16,MRBA16}. The underlying idea is that by letting a quantum simulator  work in conjunction with a classical computer, even minimal quantum resources could be made useful. In hybrid quantum-classical computation,  the computationally inexpensive calculations, which yet might  consume many qubits, are performed on a classical computer, whereas the difficult part of the computation is accomplished on a quantum simulator. The major benefit of this hybrid strategy is that it gives rise to a setup that can have   much  less stringent hardware requirements.

 

In this Letter, we propose a hybrid quantum-classical  method for solving   the quantum optimal control problem. Normally, the problem is   formulated as follows: given a quantum control system and a fitness function that measures the quality of control,    the goal is to find a control  that can achieve optimal performance. The importance of the problem lies in   its extraordinarily wide range of  applications in physics and chemistry   \cite{BCR10}.
However, current numerical  approaches   suffer from the     scalibility issue as they involve  computation of the many time propagations of the state of the controlled system, which can be infeasible on classical computers for systems of large dimensions \cite{HKCHK11}. 
To   address this computational challenge, we develop quantum versions of   gradient-based optimal control   algorithms \cite{Khaneja05}.
We show that,  given a reliable quantum simulator that    efficiently simulates the  controlled quantum evolution, then under reasonable conditions this simulator can be used to efficiently estimate both the fitness function and  its gradient.  Additionally, a classical computer is employed to   store the control parameters as well as to determine  the search direction in each iteration according to the gradient information that it receives from the simulator. Working in such a quantum-classical manner, there can be expected a significant saving of memory cost and time cost and hence an enhancement of the ability of solving the quantum optimal control problem for large-size quantum systems.
%

The proposed hybrid scheme is amenable to experimental implementation with current state-of-the-art  quantum technology. Here, we   also report a first experimental realization  of our scheme   on a   nuclear magnetic resonance (NMR) system. The  experimental results confirmed the feasiblity of our method and  show excellent performance   in obtaining high-quality optimal control solutions.

\begin{figure}[t]
\centering
\includegraphics[width=\linewidth]{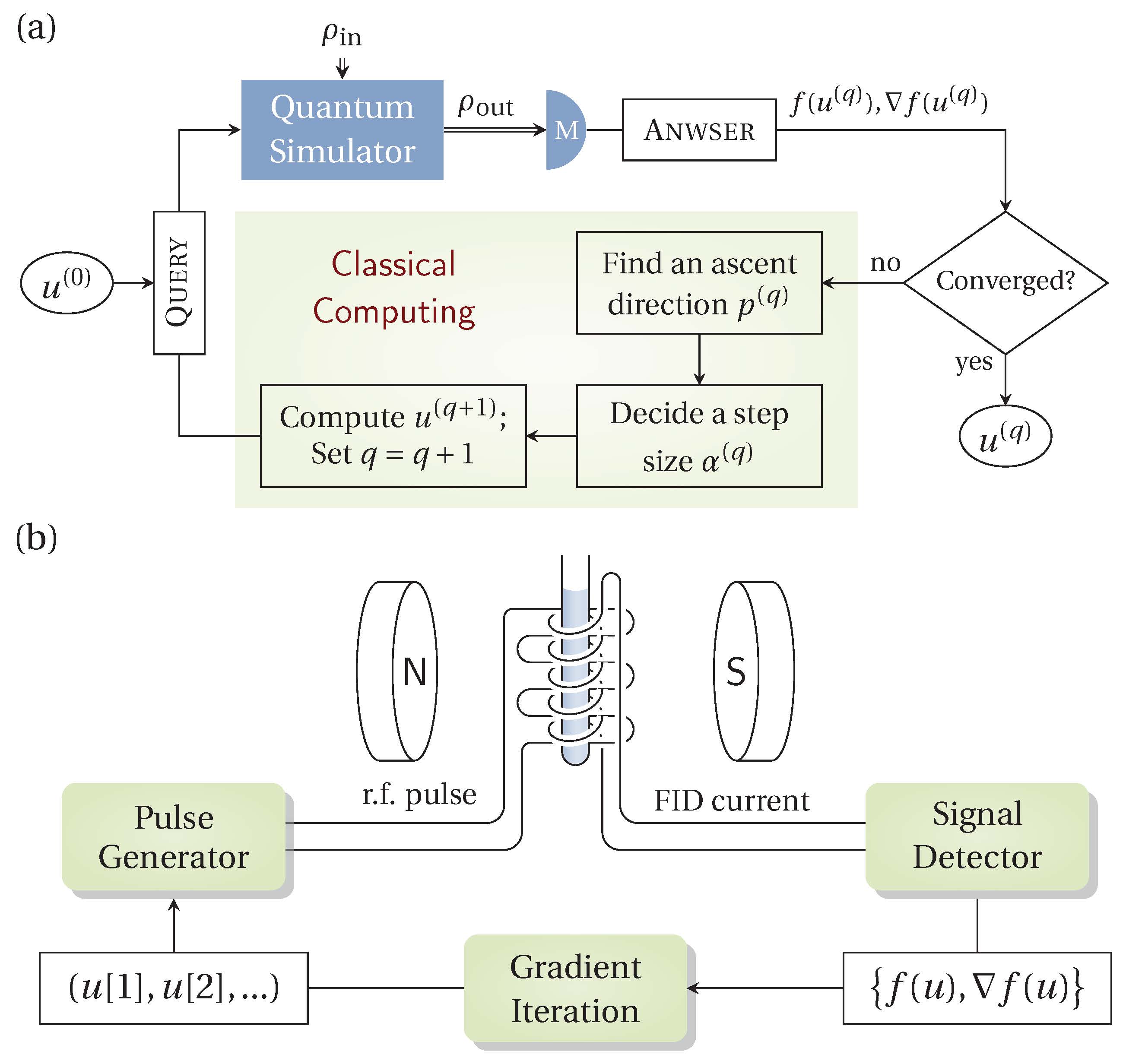}
\caption{(a)  Hybrid quantum-classical approach to gradient-based  optimal control iterative  algorithms, wherein the quantum simulator is combined with classical computing devices    to jointly implement the procedure of optimal control searching.  Here,  $\rho_\text{in}$ is input state, $\rho_\text{out}$ is output state, double-lined arrows signify quantum information, and $\rm M$ represents quantum measurement. 
(b) Schematic diagram of an NMR based  implementation of  the quantum-classical hybrid optimal control searching. The sample  consists of an ensemble of spins and serves as a quantum processor. Query  is encoded in input radio-frequency (r.f.) control pulse and the answer that the sample generates is extracted from observing the free induction decay (FID).
}
\label{oracle}
\end{figure}

\emph{Theory}.---To start, we briefly describe  the    quantum state engineering problem. Consider an $n$-spin-$1/2$ quantum spin system, which evolves under  a local Hamiltonian $H_S = \sum^L_l {H_l}$. Here each of the $L$ terms $H_l$ acts on a   subsystem containing at most a constant number of spins. Such form of Hamiltonian can be efficiently simulated  \cite{Lloyd96} and can describe a variety of quantum systems, e.g.,  quantum Ising model and   Heisenberg  model. Suppose the system is manipulated with a transverse time-varying magnetic control field $u(t) = (u_x(t), u_y(t)): t\in[0,T]$.  Let $\sigma_x$, $\sigma_y$ and $\sigma_z$ denote the Pauli  operators, then the  control Hamiltonian  reads
${H_C}(t) =  \sum\nolimits^n_{k=1} {\left( {{u_x}(t)\sigma_x^k + {u_y}(t)\sigma_y^k} \right)}$, in which  $\hbar$ is set as $1$ and the gyromagnetic ratios are not written explicitly. 
The control task is to steer the system between  states of interest in the Liouvillian space.   Normally,  we define   a fitness function to give a    performance  metric of the control. To this end, a set of operators $\mathcal{P}_n = \left\{ P_k \right\}_{k=0}^{4^n-1} = \left\{I,\sigma_x, \sigma_y, \sigma_z\right\}^{\otimes n}$, with $I$ being the $2 \times 2$ identity, is introduced. It constitutes an orthonormal basis of the state space: $\operatorname{Tr}\left( P_k P_j \right)/2^n = \delta_{kj}$ for $k, j=0, ..., 4^n-1$. Thus any state  can   be represented as a vector with respect to $\mathcal{P}_n$. Let the system's starting point be $\rho_i$ and the target be $\bar \rho = \sum\nolimits_{s\in S} {x_s P_s}$ where  $S$ is the index set for $s$.  As we are considering closed system engineering,   $\rho_i$ should be  unitarily convertible to $\bar\rho$.  
Now the state-to-state transfer task is formulated as the  quantum optimal control problem \cite{Khaneja05}:
\begin{align}
\max \quad & f(U(T){\rho _i}{U(T)^\dag },{\bar \rho})  = \operatorname{Tr}\left( {U(T){\rho _i}{U(T)^\dag } \cdot {\bar \rho}} \right)/2^n,  \nonumber \\
\text{s.t.} \quad & \dot U(t) =  - i\left[ {{H_S} + \sum\limits_{k = 1}^n { \left( u_x (t) \sigma_x^k + u_y(t) \sigma_y^k \right)} } \right]U(t),   \nonumber 
\end{align}
where $U(0) = I^{\otimes n}$ and $f$,  the fitness function, is expressed as a functional of the input control $u (t)$ and  may possess many local extrema.    
Except for  relatively small systems with two or three qubits \cite{KBG01,CHKO07},  analytically solving the problem  for generic $H_S$  is difficult.

Generally one   must   resort to numerical investigations, and the most favored approach  is to employ   gradient-based optimization methods.  
A gradient-based    algorithm   generates a sequence of iterates $u^{(0)}, u^{(1)}, ...$, which  starts from a designed trial input or even  simply a  random guess, and  stops when  a certain termination condition is fulfilled \cite{NW06}.  
The move from one   iterate $u^{(q)}$ ($q\ge 0$) to the next follows the line search strategy 
\begin{equation}
u^{(q+1)} = u^{(q)} + \alpha^{(q)} p^{(q)},
\end{equation}
that is, it  first fixes a search direction $p^{(q)}$ and then  identifies a move distance $\alpha^{(q)}$ along that direction. The computation of $p^{(q)}$ makes use of  information about $f$ and the  gradient $\nabla f$ at current iterate $u^{(q)}$, and possibly also information from earlier iterates. The step size   $\alpha^{(q)}$  is chosen such that    a sufficient increase in $f$ can be acquired. The algorithm succeeds if the sequence $f( u^{(0)}), f( u^{(1)}), ...$ converges to a desired local extremum. 
There exist    various types of gradient-based algorithms, which are classified based on the method used for determining the search direction.  For example, the known gradient  ascent pulse engineering (GRAPE) \cite{Khaneja05} algorithm  finds   local extrema by  taking steps proportional to the gradient, while conjugated gradient   \cite{BSV08}  and quasi-Newton methods \cite{FSGK11} would search along other gradient-related directions that allow for faster convergence speed.

Here we develop a hybrid quantum-classical framework for gradient-based optimal control. It would be convenient to  cast the ideas in terms of  the  standard  oracle-based optimization model   \cite{NY83,Nesterov04}. Consider an oracle function $\mathcal{O}: u  \to \left\{ f(u),  \nabla f(u) \right\}$ which, when queried at any point $u$,     gives the corresponding value of $f$ and $\nabla f$. 
Obviously that constructing such an oracle  $\mathcal{O}$     represents the most computationally resource-consuming part of the optimization procedure, and we propose     to realize it  with  using   a   reliable quantum simulator. The
simulator does not necessarily have to be universal. For instance, it can just be provided by  the controlled system itself \cite{FM15}.   The simulator  works   with a classical computer which   stores   the control variables, and if necessary     records   all   iterative information.  Our hybrid scheme consists of successive rounds of control updates, see Fig. \ref{oracle}. For each round the classical computer first sends  the current point   $u$ to the oracle $\mathcal{O}$ as   input meaning that it is posing a query,   and then  according to the answer of $\mathcal{O}$ it  executes a line search subroutine so as    to  decide  at which   point the next   query should be made.  Here, the query   is encoded in control pulse  and the answer is extracted through quantum measurements on the final state of the simulator.

So far we have not mentioned the convergence properties of the optimization. 
Gradient-based algorithms may get trapped at suboptimal  points. Yet  researches show that,  under certain conditions most of the control landscapes are trap free and convergence to an optimal solution is usually fast \cite{PT11}. In our    hybrid quantum-classical scheme, the only change is that we use  quantum resources to implement the oracle function $\mathcal{O}$. Therefore,  the   convergence properties  will remain unchanged  as long as   our  quantum simulator is sufficiently trustable.  

Now we explain   how the oracle $\mathcal{O}$ is quantumly constructed. We use the  number of experiments needed to compute $\mathcal{O}$ as a complexity measure  of the method. 
Running the numerical   optimization requires that the control field be discretized. Let the pulse $u(t)$  be divided into $M$ slices with each time slice being of constant magnitude and fixed length $\tau = T/M$. In consideration of memory cost, $M$ should be polynomially scaled.
The $m$-th slice control $u[m]$ generates the  propagator
\begin{equation}
{U_m} = \exp\left\{ - i\left[ {{H_S} + \sum\limits_{k=1}^n {\left( {u_x}[m] \sigma_x^k + {u_y}[m] {\sigma_y^k} \right)} } \right]\tau \right\}.  \nonumber
\end{equation}
For notational brevity let  $U_{m_1}^{m_2}$ denote  ${U_{m_2}}  \cdots {U_{m_1+1}} {U_{m_1}}$ where $m_2 \ge m_1$. So the final state is $\rho_f = U_1^M {\rho _i}  {U_1^M}^\dag $. We hence have the following expression for $f$
\begin{equation}
f = \operatorname{Tr}\left( { {\rho _f} }  \bar\rho  \right)/2^n = \sum\limits_{s \in S} {{x_s}\operatorname{Tr}\left( {\rho_f {P_s}} \right)/2^n}. \label{f}
\end{equation}
It can be readily seen from the equation that, rather than full  tomography of final state, $f$ can be directly measured with   $\left| S \right|$ experiments. That is, for the $s$-th experiment we first initialize our simulator at $\rho_i$, then simulate the system evolution under control $u$   and then measure the final state with   basis operator $P_s$. After this we sum up all the measurement results according to Eq. (\ref{f}) and hence obtain an estimation of $f$.

Next let us see how  to compute the $2M$-dimensional gradient vector $g= \nabla f = ({g_x}[m],{g_y}[m])$, where ${g_\alpha}[m] = \partial f/\partial {u_\alpha}[m]$ ($\alpha =x$ or $y$). To  first order approximation, it is evaluated as \cite{Khaneja05}
\begin{equation}
{g_{\alpha}}[m] =  \sum\limits_{k = 1}^n {\operatorname{Tr} \left( { - i\tau {U_{m+1}^M} \left[ {\sigma_{\alpha}^k,{U_1^m}  {\rho_i} {U_1^m}^\dag } \right] {{U_{m+1}^M}}^\dag  \bar\rho} \right)/2^n}.   
\end{equation}
The approximation is good if $\tau$ is sufficiently small. Note that for any operator $\rho$, there is
\begin{equation}
\left[ \sigma_{\alpha}^k,\rho  \right] =  i\left[ R^k_{\alpha}\left(\frac{\pi}{2}\right) \rho  {R^k_{\alpha}\left(\frac{\pi}{2}\right)}^\dag - R^k_{\alpha}\left(-\frac{\pi}{2}\right) \rho  {R^k_{\alpha}\left(-\frac{\pi}{2}\right)}^\dag  \right],   
\label{commutator}
\end{equation}
in which $R^k_{\alpha}(\pm \pi/2)$ is the $\pm \pi/2$ rotation about $\alpha$ axis on the $k$-th qubit. The essential point  is that we can compute the commutator by means of  local qubit rotations. Substituting the formula into $g$,  
\begin{equation}
g_{\alpha}[m] = \frac{\tau }{2}    \sum\limits_{k = 1}^n   \left[  \operatorname{Tr}\left( \rho_{\alpha_+}^{km} \bar\rho \right) -  \operatorname{Tr}\left( \rho_{\alpha_-}^{km} \bar\rho \right)  \right]/2^n ,
\label{g}
\end{equation}
where
$\rho_{\alpha_\pm}^{km} =   U_{m+1}^M  R_{\alpha}^k \left(\pm \pi/2 \right) {U_1^m}   {\rho _i}  \left({ { U_{m+1}^M}  R_{\alpha}^k \left(\pm \pi/2 \right) {U_1^m} }\right)^\dag$.
Therefore, to obtain the $m$-th component of $g_\alpha$, we perform $2n$ experiments: we (i)  sequentially take out an element from  the operation set $\left\{ R^k_\alpha(\pm\pi/2) \right\}_{k=1,...,n}$, and insert it  after the $m$-th slice  evolution; (ii)  measure the distances of the final states with respect to $\bar\rho$ and (iii) combine all the measurement results according to Eq. (\ref{g}).  
A quick calculation shows that in each round of iteration in total $4n M \left| S\right|$ experiments are needed to perform  gradient estimation.


Summarizing the above derivations, we conclude that in total we need to perform $(4nM+1)\left| S \right| $ experiments on the quantum simulator  to estimate $f$ and $g$. 
It is interesting to seek for instances for which  our scheme can be qualitatively advantageous over  conventional  approaches. Obviously that   target states   possessing exponential number of nonzero components require  also that  many measurements to  take. This implies that, to ensure  the whole process be feasible, we have to restrict consideration to  specific kind of target states. An important fact in quantum computing says that, to build up quantum operations  out of a small set of elementary gates is generically inefficient \cite{NC00}. In other words, there are overwhelmingly many states that are complex in the sense that they take exponential size of quantum circuit to approximate.
Therefore, it makes sense if we restrict to  relatively less complicated states, for example those which admit sparse representation with respect to some  basis, where the basis  fulfils the condition that measurement of any its element  consumes only polynomial resources. In present setting, we will be interested in $\left| S \right|$-sparse states under basis $\mathcal{P}_n$ with $\left| S \right| \ll \left| \mathcal{P}_n \right|$.   
Despite of the problem simplification,  from the practical side  they are undoubtedly still difficult tasks at current level of large-system control technology. Sparsity assumption drastically reduces the time cost  for physically implementing  $\mathcal{O}$ and in consequence  the great chance of our  oracle machine model to   provide significant speedup.

\begin{figure}[t]
\centering
\includegraphics[width=\linewidth]{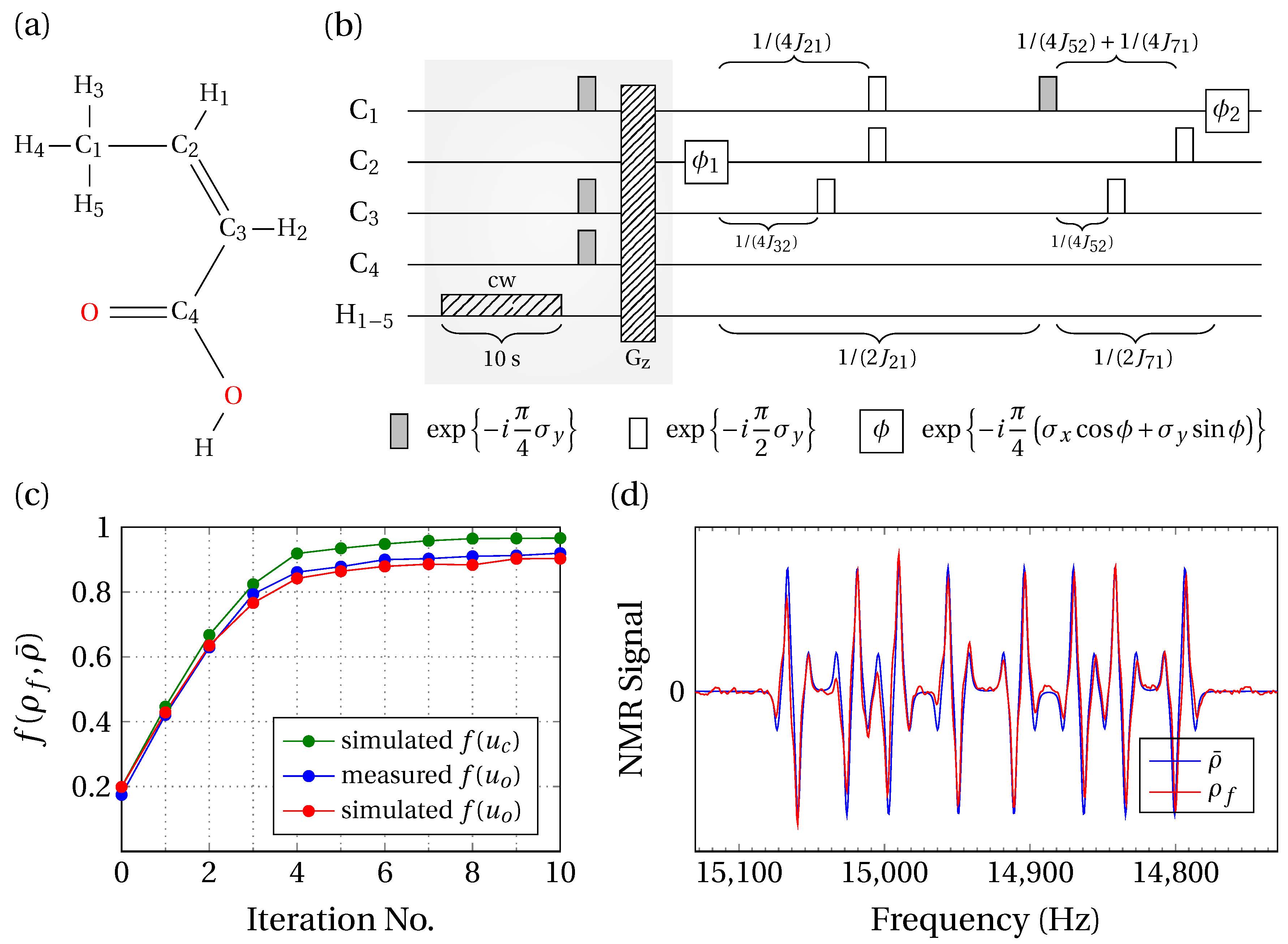}
\caption{(a) Molecular structure of    crotonic acid. (b) Pulse sequence scheme for our multiple-quantum coherence generation experiment. The grey part is designed to reset the system back into  $\rho_i$, and the preparation part is an approximate circuit (in which cw: continuous wave; G$_\text{z}$: gradient pulse along $z$ axis; $\phi_1 = -18^\circ$ and $\phi_2 = 82^\circ$) aimed for making the transform $\rho_i \to \bar \rho$. (c) Iterative results for our   system. Here $u_c$ and $u_o$ denote the controls obtained by searching on a classical computer and on the sample  respectively.  (d) NMR spectrum of $\rho_f$ after 10 times of iteration under the observation of C$_2$. It is placed  with the simulated ideal spectrum of target state $\bar\rho$ together for comparison.
}
\label{result}
\end{figure}

\emph{Experiment}.---We chose the fully  $^{13}$C-labeled crotonic acid  as our test system, on which we demonstrated the idea of using the sample   to compute its own optimal control pulse.    Fig. \ref{result}(a) shows the molecular structure of the sample. The four carbon nuclei   plus the five proton nuclei constitute a nine-spin system, in which the  methyl protons H$_3$, H$_4$ and H$_5$ are  chemically and magnetically equivalent and hence are indistinguishable. Experiment is carried on a Bruker Avance III 400 MHz  spectrometer at room temperature. The system  Hamiltonian takes the form: 
${H_S} = \sum\nolimits^n_{k =1} {{\Omega _k}\sigma_z^k/2}  + \pi \sum\nolimits^n_{k<j} {{J_{kj}} \sigma_z^k \sigma_z^j/2}$, where $\Omega_k$ is the precession frequency of the $k$-th spin, and $J_{kj}$ is the coupling between the $k$-th and $j$-th spin, see supplementary material   \cite{S} for their values. To describe states of the nuclei, we use deviation density matrices, that is,  the traceless part of the density matrices up to an overall scale \cite{SELBE83}.
Our goal is to create the seven-correlated state $\bar \rho = \sigma_z^1 \sigma_z^2 \sigma_z^3 \sigma_z^5 \sigma_z^7 \sigma_z^8 \sigma_z^9$, which is the largest multiple-correlated  operator that can be directly observed from the spectrum.  Observation is made on  C$_2$ because all the couplings  are adequately resolved.  Our experiment   is divided into two parts: reset and preparation.

In the reset part we rest the system to a fixed initial state $\rho_i$, which has to be unitarily equivalent to $\bar \rho$. So the system's equilibrium state is not considered  because it has different spectra with that of $\bar \rho$. Although there are many candidates, we choose $\rho_i \propto \sigma_z^2$ for convenience of observation and design a corresponding initialization procedure, see Fig. \ref{result}(b). First, we apply a  continuous wave (cw)  on the proton channel. Because of the steady state hetero-nuclear Overhauser effect (NOE) \cite{L08}, provided that the cw irradiation is  sufficiently long and strong,  then the system will be  driven asymptotically into a steady state $\rho_{ss}$ of the form: $\rho_{ss} = \sum\nolimits_{k=1}^4 {\epsilon^k_{ss} \sigma_z^k}$, that is,  the  protons are saturated. In experiment, the irradiation is set to be 10 s of duration and 2500 Hz of magnitude. As expected we see the establishment of the steady state, in which only the carbons' polarizations are left, but with enhanced  bias compared to the equilibrium state. For example, the boost factor of C$_2$ is about 1.8. Next, we  retain just the signal of   C$_2$  by first rotate the polarizations of other carbons to the transverse plane and then destroy them with $z$ axis gradient field.  This gives the desired initial state $\rho_i$.

The preparation part aims to  steer   $\rho_i$ towards $\bar\rho$. To give a good initial  control guess to accelerate convergence, we designed an approximate preparation circuit. The approximate circuit is constructed based on a simplified system Hamiltonian which ignores the small couplings and  the small differences between large couplings of the original Hamiltonian. Such simplification manifests   which   couplings are allowed to evolve for preparing $\bar \rho$ thus enables  direct circuit construction, see Fig. \ref{result}(b). The circuit thus constructed, if we turn back to the real Hamiltonian, generates a final state that deviates $\bar\rho$ only slightly: $f \approx 0.9824$. Moreover, the circuit length is 16.36 ms, much shorter than system's relaxation time, so the preparation stage can be  taken as unitary.  In order that the number of control parameters  after pulse discretization be as few as possible, we adopt relatively large time step length $\tau = 20$ $\mu$s. We   use Gaussian shaped  selective pulses to implement the rotational gates.  Each  selective pulse has  its pulse-width determined according to which qubit  it is acting on. Excluding the free $J$ evolutions, we have  in total $2 \times 108$ nonzero pulse parameters  to be optimized. We have  employed a compilation procedure   \cite{RNLKL08, LCLP16}  to systematically reduce the errors that come in when the ideal rotational operations are implemented through soft selective pulses, yet   $f$ still drops severely.    
Therefore,  some extent of  pulse optimization is necessary.  

We add a small amount of random disturbances  to  the above constructed selective pulse network. The purpose of doing so is to  start the oracle iteration from a relatively low-quality  control and hence to witness a more notable  rising  of $f$.   According to our previous  analysis, we  roughly figure out the experiment time cost for each round of iteration to be about $5$ hours.   We have demonstrated the query action on the sample  for $10$ times.  Fig. \ref{result}(c-d) shows the experimental results, from which we see clearly that     the successively updated pulse   is indeed approaching a solution of the optimal control problem.
Because that   measurement inaccuracies induce errors in   gradient estimation,  it is as expected that some degree of deviation of the experimental growth of $f$     from that performed on a classical computer appears. Therefore, the important  challenge left open is to understand quantitatively that how measurement inaccuracies affect the convergence efficiency. 

\emph{Discussion}.---From the control theory perspective, the apparatus in our experiment,   including a control input generator, a sample of molecules and a measurement device, interact as a closed learning loop. In each cycle of the loop, the fitness information learned from the sample  directs the  optimization to achieve a given control objective. Such strategy has the advantage of reliability and robustness. Learning algorithm is the crucial ingredient, and previous  studies have been mainly focused on using stochastic searching strategies such as evolutionary algorithms \cite{JR92,Rabitz00}. We here have shown that a large class of gradient-based methods    can also be incorporated into the closed loop learning control model.  This will be important for realizing  high-fidelity quantum control experiments, such as is   needed in the fields of    quantum information processing and spectroscopy.

Future work will  seek to gain a better understanding of the feasibility of    the hybrid quantum-classical approach to quantum optimal control. NMR is an excellent platform on which to test various quantum control methods, but for our scheme its drawback is  the relatively long reset (relaxation) time. It can be envisioned that  on some other quantum information processing candidate systems that are with much shorter operation time and relaxation time \cite{LJLNMO10}, the search  process  may get several orders of magnitude faster. We expect the methodology developed in this work can promote studies of scalable quantum controls on larger quantum systems.

This work is supported by the  National Basic Research Program of China (973 Program, Grant No. 2014CB921403), the National Key Research and Development Program (Grant No. 2016YFA0301201), the Foundation for Innovative Research Groups of the National Natural Science Foundation of China (Grant No. 11421063),  the  Major Program of the National Natural Science Foundation of China (Grant No. 11534002), the State Key Development Program for Basic Research of China (Grant Nos. 2014CB848700 and 2013CB921800), the National Science Fund for Distinguished Young Scholars (Grant No. 11425523), and the National Natural Science Foundation of China (Grant No. 11375167 and Grant No. 11605005).

\newpage

\begin{widetext}
\section{Supplementary Material:\\  ``Hybrid Quantum-Classical Approach to Quantum Optimal Control"}
\end{widetext}

\section{Gradient-based Methods} 
Our aim is to  develop a hybrid quantum-classical framework for gradient-based optimal control. As we have described in the main text, a gradient-based iterative    algorithm generates successive iterates from an initial guess $u^{(0)}$, where the move from one iterate $u^{(q)}$ to the next $u^{(q+1)}$ consists of the following steps:  
\begin{enumerate} 
\item[(1)] compute $\mathcal{O}: u^{(q)} \to \{f(u^{(q)}), \nabla f(u^{(q)}) \}$,  
\item[(2)] determine a gradient-related search direction $p^{(q)}$,  
\item[(3)] find an optimal step size \begin{equation} \alpha^{(q)} = \arg \max_{\alpha} f(u^{(q)} + \alpha p^{(q)}), \nonumber \end{equation} 
\item[(4)] update $u^{(q+1)} = u^{(q)} + \alpha^{(q)} p^{(q)}$.  
\end{enumerate} 
We have shown that  step (1) can be done with using a quantum simulator. Step (3) is  a one-dimensional search. Often, the step size need not be determined exactly. Inexact   search   may be performed in a number of ways, such as a backtracking line search or using the Wolfe conditions. They   can be done by a classical computer. Now we take a closer look at how step (2) is accomplished using the hybrid quantum-classical setup.

Different types of gradient-based algorithms use  different strategies to compute the search direction. Representative  gradient-based algorithms include  gradient ascent, conjugate gradient and quasi-Newton methods \cite{NW06S}. 

\emph{Gradient ascent}. The gradient ascent method simply takes  steps along the gradient.
\begin{equation} 
p^{(q)} = \nabla f(u^{(q)}).  \label{G}
\end{equation}

\emph{Nonlinear conjugate gradient ascent}. In numerical optimization, nonlinear conjugate gradient method is a widely used method for solving nonlinear optimization problems. The search direction is determined by
\begin{equation} 
p^{(q)} = \nabla f(u^{(q)}) - \beta^{(q)} p^{(q-1)}, \label{CG}
\end{equation}
where $\beta^{(q)}$ is computed according to  Fletcher–-Reeves formula
\begin{equation} 
\beta^{(q)} = \frac{ ( \nabla f(u^{(q)})  )^T \nabla f(u^{(q)}) }{  ( \nabla f(u^{(q-1)})  )^T \nabla f(u^{(q-1)}) }, \nonumber
\end{equation}
or Polak–-Ribi{\`e}re formula: 
\begin{equation} 
\beta^{(q)} = \frac{ ( \nabla f(u^{(q)})  )^T (\nabla f(u^{(q)}) - \nabla f(u^{(q-1)})) }{  ( \nabla f(u^{(q-1)})  )^T \nabla f(u^{(q-1)}) }, \nonumber
\end{equation}
or some other formulas such as Hestenes-Stiefel formula.

\emph{Quasi-Newton method}. Quasi-Newton methods also   require only the gradient of the fitness function to be supplied. They are able to produce superlinear convergence by utilizing information about the changes in gradients. The search direction is determined by
\begin{equation} 
p^{(q)} = H^{(q)} \nabla f(u^{(q)}), \label{QN}
\end{equation}
where $H^{(q)}$ is an approximation to the Hessian, and is also updated in each iteration.   There exist a number of different methods for   updating   $H^{(q)}$ \cite{NW06S}. For example, the  Davidon–-Fletcher–-Powell   formula  goes
\begin{equation} 
H^{(q+1)} = H^{(q)} + \frac{\Delta u^{(q)} (\Delta u^{(q)})^T}{(\Delta u^{(q)})^T y^{(q)}} - \frac{H^{(q)} y^{(q)} (y^{(q)})^T H^{(q)}}{(y^{(q)})^T H^{(q)} y^{(q)}}, \nonumber
\end{equation}
and the Broyden–-Fletcher–-Goldfarb–-Shanno formula goes
\begin{align} 
H^{(q+1)} = {} & \left(I - \frac{\Delta u^{(q)} (y^{(q)})^T}{(y^{(q)})^T \Delta u^{(q)}}\right) H^{(q)} \left(I - \frac{y^{(q)} (\Delta u^{(q)})^T }{(y^{(q)})^T \Delta u^{(q)}}\right)  \nonumber \\
{} & + \frac{\Delta u^{(q)} (\Delta u^{(q)})^T}{(y^{(q)})^T \Delta u^{(q)}}, \nonumber
\end{align}
where $\Delta u^{(q)} = u^{(q+1)} - u^{(q)}$, $y^{(q)} = \nabla f(u^{(q+1)}) - \nabla f(u^{(q)})$ and the initial   $H^{(0)}$ can be set as the identity matrix.  

We can see from Eqs. (\ref{G}--\ref{QN}) that, in their quantum-classical hybrid implementation, the classical computer  should have recorded all iterative information. In computing the search direction $p^{(q)}$, the classical computer would use   information about $f$ and  $\nabla f$ at current iterate, and possibly also information from earlier iterates.

\section{Experiment} 
\begin{figure}[t]  
\centering
\includegraphics[width=\linewidth]{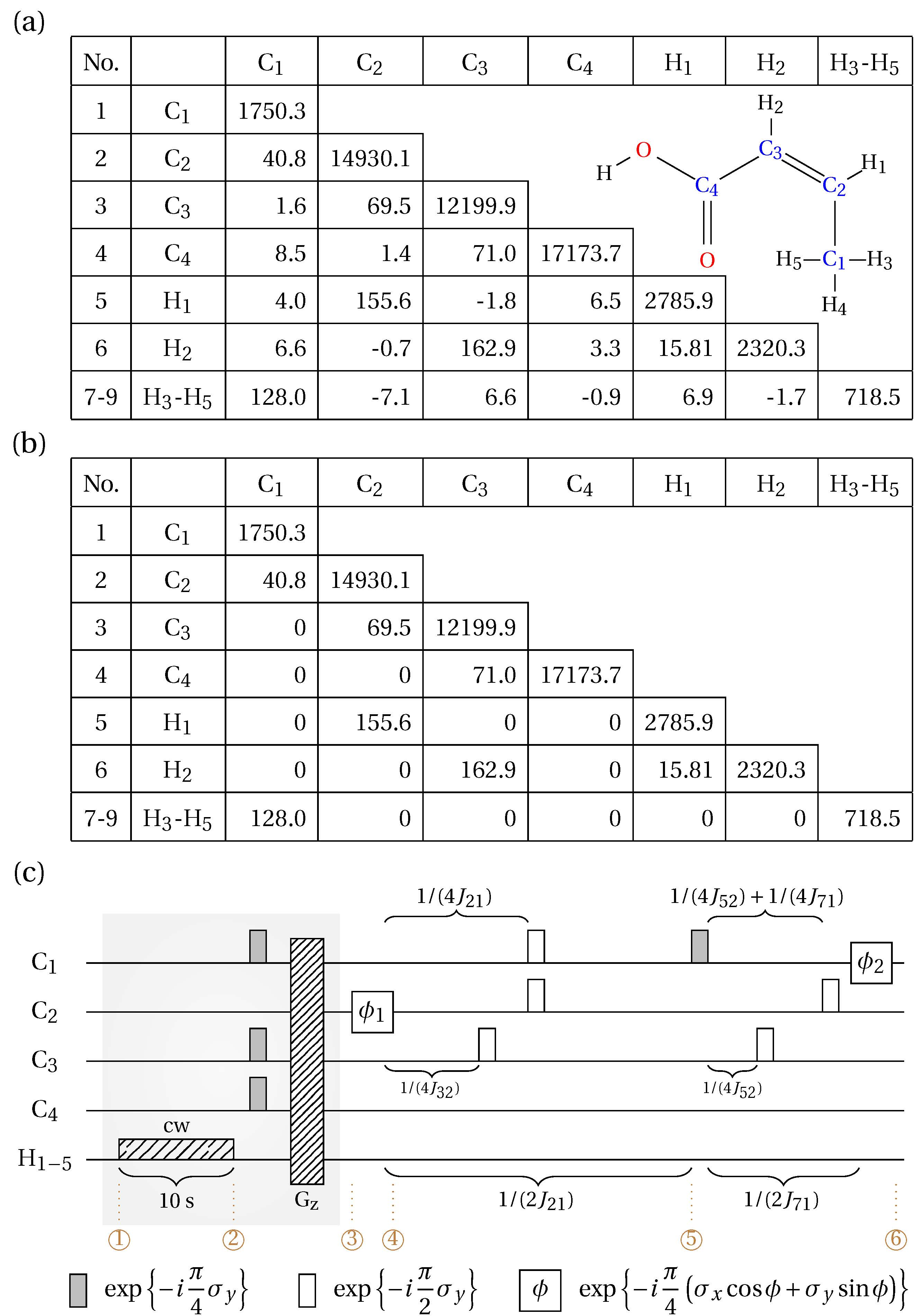}
\caption{(a) The Hamiltonian parameters (all in Hz) are experimentally determined on a 400 MHz spectrometer. In the table, diagonal elements give chemical shifts with respect to the base frequency for carbon and proton transmitters; off-diagonal elements give $J$ coupling terms. (b) Simplified coupling network. It is  obtained by ignoring the small couplings and  the small differences between large couplings of the original Hamiltonian. (c) Operation sequence for our state-to-state optimal control experiment.}
\label{circuit}
\end{figure}

In this section, we give more details of how our experiment is performed. Our experimental system is the fully  $^{13}$C-labeled crotonic acid. The system parameters are given in Fig. \ref{circuit}(a).
What we want to demonstrate in experiment is the process of using this sample to compute its own optimal control pulse. Concretely, we choose to study the following state-to-state transfer    task: 
\begin{equation}
\rho_i = \sigma_z^2 \to \bar \rho = \sigma_z^1 \sigma_z^2 \sigma_z^3 \sigma_z^5 \sigma_z^7 \sigma_z^8 \sigma_z^9.
\end{equation}
As we have described in the main text, the implementation of the hybrid quantum-classical   algorithm requires repeatedly querying the sample. Each query action  is composed of three stages: initialization, preparation and detection. The final detection of  $\bar\rho$ is made by    observing the spectrum of  C$_2$, see Fig.  for the ideal spectrum.  We shall not consider errors present in the initialization and detection step, nor the relaxation effects during the preparation step.

\subsection{Initialization}
The initialization step aims to create the initial state $\rho_i$ from the equilibrium state
\begin{equation}
\rho_{eq} = \sum_{k=1}^4 {\epsilon_{eq}\sigma_z^k} + 4 \sum_{k=5}^9 {\epsilon_{eq}\sigma_z^k} \to \rho_i = \sigma_z^2,
\end{equation}
where $\epsilon_{eq}$ is the    equilibrium polarization magnitude of carbon.
This is accomplished through a non-unitary process, which utilizes the steady state hetero-nuclear Overhauser effect as explained in the main text. The system at \textcircled{2} is $\rho_{ss} = \sum_{k=1}^4 {\epsilon_{ss}^k\sigma_z^k}$ where $\epsilon_{ss}^k$ denotes the steady state polarization magnitude. The system at \textcircled{3} is $\rho_{i} = \epsilon_{ss}^2 \sigma_z^k$. From experiment, we  have obtained that $\epsilon_{ss}^2 \approx 1.8 \epsilon_{eq}^2$. For convenience, from here on we shall rescale $\epsilon_{ss}^2$ as 1.

Note that there  is a more familiar way of producing pure operator $\sigma^2_z$, namely selectively rotate all the spins except C$_2$ and followed a longitudinal gradient pulse. However, our method here has  two important advantages, both attributed to NOE: (i) higher polarization of initial state, hence an increased signal-to-noise ratio of the  spectrum; (ii) faster reset time, NOE takes much less time to reset the system, that for our sample this is found to be about six times faster. 
Therefore,  NOE based reset procedure greatly reduces the total time cost of our experiment. 

\subsection{Preparation}

\begin{figure*}
\begin{center}
\includegraphics[width=0.85\linewidth]{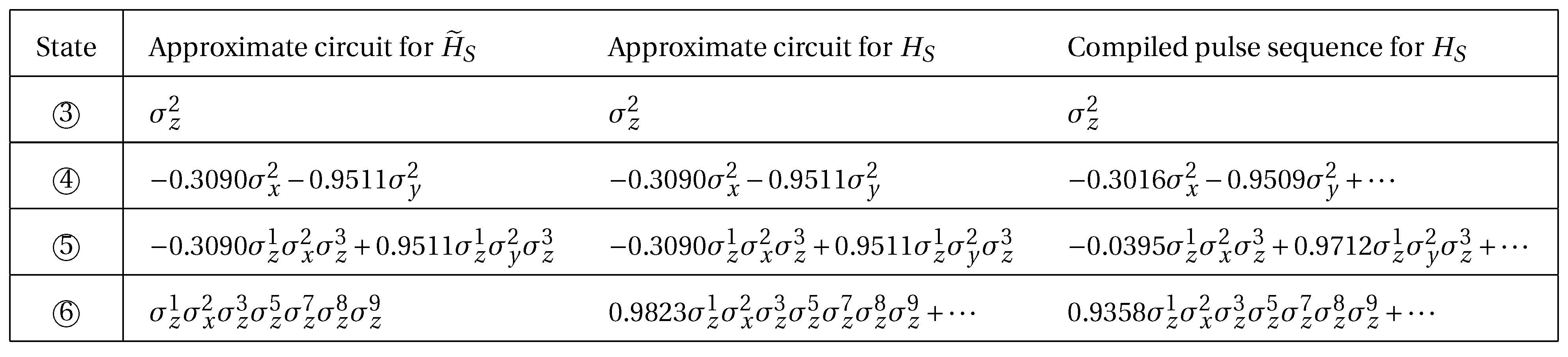}
\caption{State evolution of the quantum system under the constructed quantum circuit and the compiled selective pulse sequence.}
\label{preparation}
\end{center}
\end{figure*}

\emph{Constructing an approximate circuit}.
Firstly, we construct   an ideal preparation circuit for an approximate Hamiltonian, which of course would be an approximate circuit for the real Hamiltonian.  Consider a simplified  coupling network    as given in   Fig. \ref{circuit}(b).
Let us denote the     approximate  Hamiltonian out of this simplified  coupling network as $\widetilde H_S$. Based on the simplified Hamiltonian, we build up an ideal circuit that is able to realize the desired state-to-state transfer with perfect fidelity, see  Fig. \ref{circuit}(c).
The   circuit      proceeds in two sub-steps: in the first sub-step, we first turn $\sigma_z^2$ into 1-coherence and then generate three-correlated operators $\sigma_z^1 \sigma_x^2 \sigma_z^3$ and  $\sigma_z^1 \sigma_y^2 \sigma_z^3$ through coupled evolutions of $J_{12}$ and $J_{23}$; in the second sub-step, we first add a $\pi/2$ rotation about $y$ axis to the first spin and then generate seven-correlated operator $\sigma_z^1 \sigma_x^2 \sigma_z^3 \sigma_z^5 \sigma_z^7 \sigma_z^8 \sigma_z^9$ through coupled evolutions of $J_{17}$, $J_{18}$, $J_{19}$ and $J_{25}$. Here, in each sub-step, the   unwanted $J$ coupling evolutions should be appropriately refocused. 
Back to the real Hamiltonian $H_S$, this   circuit   generates  a final state with high fidelity $f = 0.9824$, see Fig. \ref{preparation}.

\emph{Selective pulse sequence compilation}. Next,  we seek to construct a pulse sequence that can implement the approximate circuit. 
The circuit is composed of multiple-spin rotational gates and free evolutions. To realize the rotational gates, we   take   use of frequency selective pulses.  Selective pulses   have the property of selectively exciting spins over a limited frequency region, while minimizing influences to spins that are outside this region.  For example, a rotational gate on a specific spin can be
realized by a rotating Gaussian that is on resonance with that spin.  
The approximate circuit contains in total 7 multiple-spin rotations, each implemented through a selective pulse  with the basic Gaussian selective pulse shape. The time lengths of these  selective pulses are not all the same, actually they are determined according to the frequency distances of the spins to be excited with the other spins.

In order that the number of control parameters after pulse discretization be as few as possible, we adopt relatively large time step length $\tau=20 \mu$s. Accordingly the time lengths of the selective pulses  and the free evolutions of the pulse sequence are not exactly the same as those in the circuit of Fig. \ref{circuit}(c), and they are redefined as such: $t \to \tau\operatorname{Round}(t/\tau)$ where $t$ represents any of   $1/(4 J_{21}), 1/(4 J_{32}), 1/(2 J_{21}),...$.
The     resulting pulse is of length 16.36 ms, discretized into 818 time slices and of which in total 108 slices are nonzero. Therefore, there are $2 \times 108$ nonzero pulse parameters to optimize. 

It is important to be aware of that a selective pulse just approximately implements the target operation. Various types of errors arise when  transferring a circuit directly into a selective pulse sequence without correction. And as the number of gates contained in the circuit grows large, the error accumulation will become increasingly serious. To address this problem, Refs. \cite{RNLKL08S,LCLP16S} put forward a   pulse sequence compilation program. The compilation program   systematically adjusts  the   pulse parameters of an arbitrary input selective pulse sequence so that  errors up to first-order can be corrected. The compilation procedure is efficient. With application of the  compilation method to our pulse sequence,   the control accuracy is   greatly improved, see the compiled results shown  in Fig. \ref{preparation}.  Although the compilation program can not eliminate all control imperfections    that higher-order errors still exist,  it is quite useful in  that, the    pulse sequence after compilation is of relatively high fidelity and can be used as a good starting point for subsequent gradient-based optimization. We refer the reader to   Ref. \cite{RNLKL08S} for a detailed   elaboration of how to take both pulse compilation method and optimal control method together as a basis to   pulse   design   in   large-sized quantum system control.

\emph{Time cost of the oracle function $\mathcal{O}$}.
As we have derived in the main text, it   requires $(4nM+1)\left| S \right|$ experiments to  be performed on the sample to calculate the function $\mathcal{O}: u \to (f(u), g(u))$. The time cost of each experiment is composed of three parts: (i) the time cost of   initialization, which we now denote by $T_0$; (ii) the time cost of running the circuit, which is 16.36 ms; (iii) the time cost of detection, which takes 1$\sim$2 seconds.    

For our NMR experiments, we could let the sample freely relax  back to its equilibrium state   $\rho_i$, which takes about 5 times the relaxation characteristic time of the spins, i.e., about 60 s. However,   our initialization is   performed in another approach as is described in the main text, and in doing so, the reset process could be six times faster. Therefore, our experimental time cost of the oracle function $\mathcal{O}$ is reduced to 5 hours.

\end{document}